\documentclass[final,5p,times,twocolumn]{elsarticle}

\usepackage{graphicx}
\usepackage{amssymb}
\usepackage{amsmath,bm}
\usepackage{todonotes}
\usepackage{booktabs}
\usepackage{array}
\usepackage{caption}
\usepackage{url}

\usepackage{lineno}
\usepackage{color}
\usepackage{qcircuit}

\usepackage{tikz}
\usetikzlibrary{shapes.geometric, arrows}

\tikzstyle{startstop} = [rectangle, rounded corners, minimum width=3cm, minimum height=1cm,text centered, draw=black, fill=yellow!30]
\tikzstyle{process} = [rectangle, minimum width=3cm, minimum height=1cm, text centered, draw=black, fill=white]
\tikzstyle{arrow} = [thick,->,>=stealth]

\journal{Journal Name}

\begin{document}

\begin{frontmatter}


\title{Comparative study of the ans\"atze in quantum language models }




\author{Jordi Del Castillo} 
\author{Dan Zhao}
\author{Zongrui Pei} 
\affiliation{New York University, New York, NY 10003, United States}
\ead{zp2137@nyu.edu;peizongrui@gmail.com} 

\begin{abstract}

Quantum language models are the alternative to classical language models, which borrow concepts and methods from quantum machine learning and computational linguistics. While several quantum natural language processing (QNLP) methods and frameworks exist for text classification and generation, there is a lack of systematic study to compare the performance across various ans\"atze, in terms of their hyperparameters and classical and quantum methods to implement them. Here, we evaluate the performance of quantum natural language processing models based on these ans\"atze at different levels in text classification tasks. We perform a comparative study and optimize the QNLP models by fine-tuning several critical hyperparameters. Our results demonstrate how the balance between simplification and expressivity affects model performance. This study provides extensive data to improve our understanding of QNLP models and opens the possibility of developing better QNLP algorithms.


\end{abstract}


\end{frontmatter}



\section{Introduction}
Quantum natural language processing (QNLP) is a language processing technology at the intersection of quantum computing, machine learning, and computational linguistics \cite{guarasci2022quantum,meichanetzidis2020quantum,karamlou2022quantum}. As a ``baby technology", QNLP borrows concepts and methods to provide a more efficient process for performing general text classification and generation tasks. Its goals and applications are similar to those of classical language models but with an expectation of higher efficiency and speed, assisted by quantum technology.

In recent years, large language models (LLMs) have been developed and represented by GPT, Llama, Gemini, DeepSeek and a few others. LLMs are a paradigm-shifting technology that has changed many aspects of daily life and research domains. For example, it can create short movies \cite{zhu2023moviefactory}, provide medical suggestions \cite{thirunavukarasu2023large}, analyze financial statements \cite{kim2024financial}, write codes \cite{chen2021evaluating}, and design new materials \cite{pei2024designing,pei2024towards,pei2024computer}. As its name indicates, LLMs are characterized by their tremendous amount of hyperparameters, currently as large as several trillion. The present trend of LLMs is to adopt increasingly more hyperparameters to make the models more capable, which requires more electric power and expensive graphical processing units (GPUs) to train. This trend is not sustainable, and we need new technology to maintain the expected growth of LLMs.

Quantum computing is one of the state-of-the-art research areas in computer science, applied mathematics, quantum information science, and physics. Its physical realization is a quantum circuit or a quantum computer, which consists of quantum bits or qubits. The state of a qubit is a continuous number, unlike classical bits that have only values of 0 and 1. Quantum acceleration and quantum supremacy come partly from quantum entanglement and partly from the exponential representation power of quantum computers, which allows them to represent or store large amounts of data with a few qubits \cite{arute2019quantum,boixo2018characterizing,Acharya2024}. It has the potential to revolutionize many areas, mainly due to the exponential speedup promised by the technology. It is natural to research its opportunities in NLP and language models.

One representative quantum NLP method is DisCoCat (DIStributional COmpositional CATegorical), a compositional model proposed by Coecke {\it et al.} \cite{coecke2010mathematical}. Unlike bags of words or other statistics-based mathematical models, the model can generate a string diagram for each sentence according to grammar relationship. The grammar structure is reserved by the pregroup grammar, a computational algebraic approach developed by Lambek \cite{lambek2008word}. The string diagram is then mapped into a quantum circuit according to some ans\"atze or parametrization scheme. Different ans\"atze are suitable for different problems. This is like the kernels in machine learning. For linear problems, linear kernels perform better than non-linear ones. One must perform tests to find the best ans\"atze for a specific problem. The original string diagram or pregroup diagram usually involves many qubits and suffers from the risk of overparametrization for small datasets; therefore, some simplification is used, like the removal of ``cups". Here, a cup is illustrated as a ``U" shape in the diagram, equivalent to a summation over all basis vectors $\sum_i \langle ii |$. After simplification of the diagram, the quantum circuit involves fewer qubits and quantum gates. We will provide more details in the methodology part.


A few quantum versions of the classical language models have been developed, such as the so-called quantum self-attention neural networks for text classification \cite{li2024quantum} and the quantum transformer model \cite{khatri2024quixer}.
Another noteworthy algorithm is the quantum counterpart of the ``bag of words" method for text classification \cite{lorenz2023qnlp}. The core idea is to treat each word as a square matrix $M_i$ and define a mapping $m: \{M_i\} \rightarrow V$. After matrix multiplication, the output matrix is the same size as the input matrices, i.e., $V=M_1\cdot M_2 \cdot M_3 \dots =\prod_i M_i$. These exciting progresses signal the promising future of QNLP.


Many challenges must be solved before QNLP becomes practically applicable, as with its classical counterpart. For example, (i) it is unknown how to extend current QNLP methods to treat arbitrary long sentences; (ii) more effort is needed to explore efficient and user-friendly ways to transform sentences into a format that can run in quantum hardware; (iii) there is a lack of systematic methods to maximize quantum advantages (by hyperparameter fine-tuning, new ans\"atze and framework designing).
As a collective effort towards overcoming these challenges, we will systematically study the existing ans\"atze and their performance. Our comparative study of hyperparameters, such as the number of qubits and the depth of quantum circuits, is a significant stride towards understanding the influence of these parameters on QNLP. This understanding paves the way for the future design of new ans\"atze.




\section{Theoretical background} 

\subsection{Workflow}
\begin{figure}[h]
        \centering
        \begin{tikzpicture}[node distance=2cm]
            \node (sentence) [startstop] {Sentence};
            \node (stringdiagram) [startstop, below of=sentence] {String diagram};
            \node (rewritten) [startstop, below of=stringdiagram] {Rewritten string diagram};
            \node (quantumcircuit) [startstop, below of=rewritten] {Quantum circuit / tensor net};
            \node (applications) [startstop, below of=quantumcircuit] {Applications};

            \draw [arrow] (sentence) -- node[right] {Parsing/encoding} (stringdiagram);
            \draw [arrow] (stringdiagram) -- node[right] {Rewriting} (rewritten);
            \draw [arrow] (rewritten) -- node[right] {Parametrization/Ansatz} (quantumcircuit);
            \draw [arrow] (quantumcircuit) -- node[right] {Training/optimization} (applications);
        \end{tikzpicture}
    \caption{The workflow of mapping a sentence onto a quantum circuit and optimization. The workflow consists of four steps. The first three steps transform sentences into quantum circuits, while the last step is model training and parameter optimization. The first step is to parse the sentence into a syntax tree and encode it in a string diagram according to the pregroup grammar. The string diagram is simplified by rewriting. This step is optional, mainly to reduce the complexity of the diagram and the number of parameters involved later in the quantum circuit. The rewritten diagram is parameterized by an ansatz and transformed into a quantum circuit or a tensor net. The main difference between quantum-circuit and tensor-net representation is whether tensors or quantum gates represent the words. Finally, the parametrized model is optimized by training on input data. After optimization, the model is ready for various task-specific applications.}
    \label{fig:sentence_to_qc}
\end{figure}
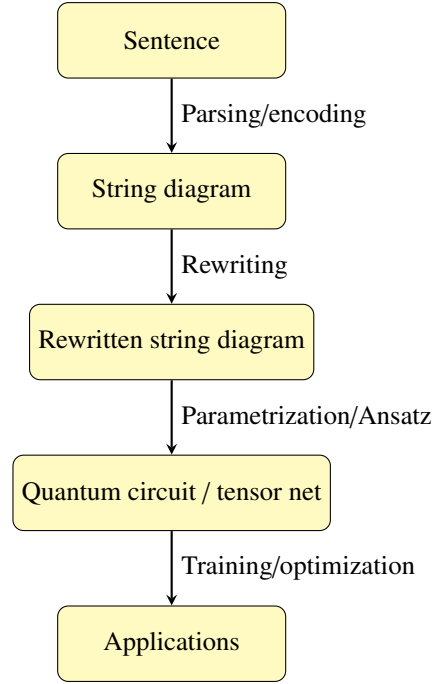

\begin{figure*}
    \centering
    \includegraphics[width=1.0\linewidth]{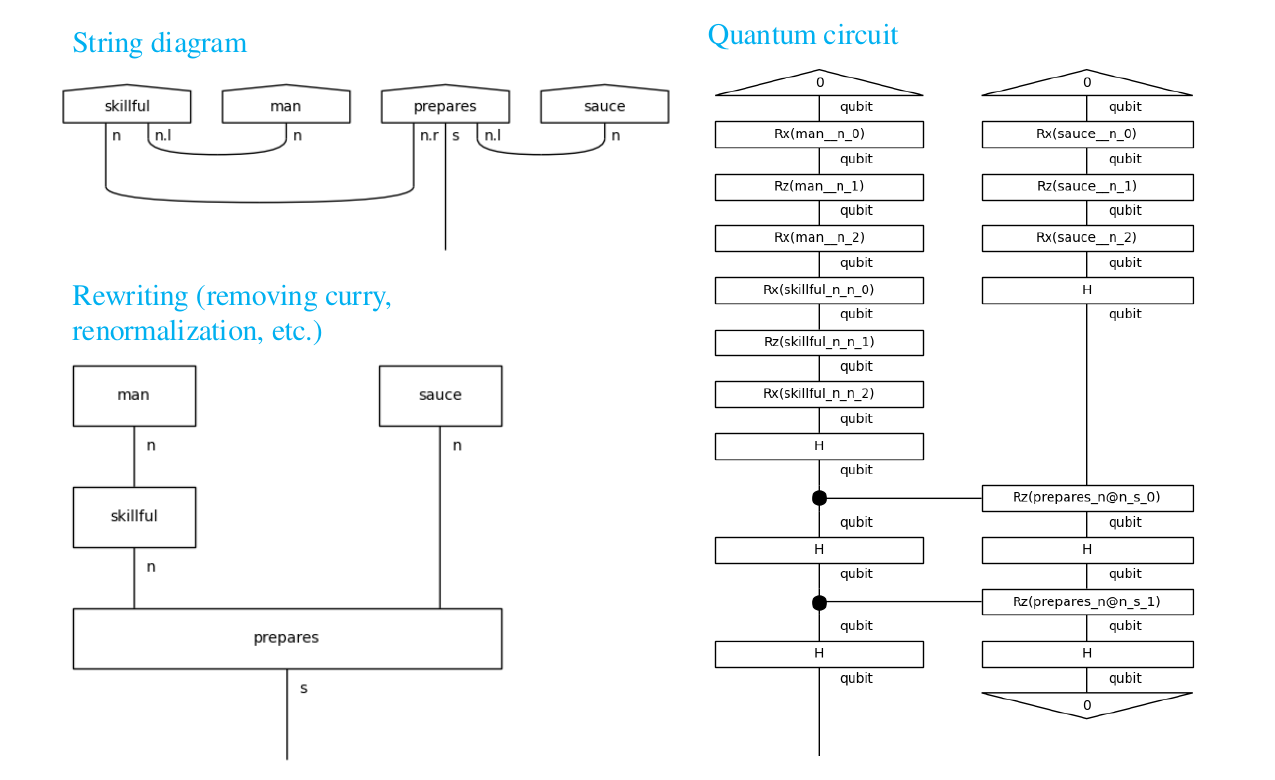}
    \caption{Example of mapping a sentence onto a quantum circuit. The string diagram, rewritten string diagram and quantum circuit are shown, taking the sentence ``skill man prepares sauce" as an example.}
    \label{fig:qnlp-mapping}
\end{figure*}

The general training procedure of QNLP consists of preprocessing data, encoding data into the quantum space, choosing the optimization algorithms, setting up the hyperparameters (like the number of shots and the depth of quantum circuits), etc. Data preprocessing includes removing punctuation and special symbols and stopping words such as ``a"and ``an". Currently, QNLP algorithms have limited capability and take sentences of the same length.
The workflow of this study starts with encoding the sentences into the quantum space (Figure \ref{fig:sentence_to_qc}). An example is illustrated in Figure \ref{fig:qnlp-mapping}. The procedure consists of four steps: (i) use pregroup grammar to represent a sentence using a diagram; (ii) some rewriting is used to simplify the figure, such as removal of cups and curry. This step is optional but beneficial, reducing the number of parameters in the model and improving the efficiency and accuracy of the model; (iii) choose an ansatz (like IQP; details are described later) and then use modules like PennyLanne/Tket to transform the ansatz into a specific quantum circuit. Different ans\"atze have different numbers of parameters and various combinations of quantum gates. The ansatz is a critical factor in determining the performance of QNLP models. (iv) The fourth step is to optimize the model by training on input data. After these four steps, the optimized model is ready for various applications.

\subsection{Pregroup grammar, quantum circuits, and tensors}
This study adopts the distributional compositional semantics (DisCo) model that includes both the probability feature of modern language models (like a bag of words) indicated by ``distributional" and the grammatical structure indicated by ``compositional". Compositional semantics obtains the meaning of a sentence from component words, their types, and the grammar structures.
The grammatical structure is preserved by a computational linguistic method, i.e., the pregroup grammar, which assigns a ``type" to each word. Types are defined in such a way that when neighboring words in a sentence can perform matrix multiplications, which results in an identity matrix and disappears. For example, a noun is defined in a vector space $\mathcal{N}$, an adjective in a space $\mathcal{N} \otimes \mathcal{N}$, and a transitive verb in a space $\mathcal{N} \otimes \mathcal{S} \otimes \mathcal{N}$.  A vector $\bm{p}$ has both left and right adjoints denoted by $\bm{p}^l$ and $\bm{p}^r$ respectively. They follow the reduction rules:
\begin{equation}
\bm{p} \cdot \bm{p}^r= \bm{1}; ~~~ \bm{p}^l \cdot \bm{p} =\bm{1}.
\end{equation}

The type of transitive verb is $\bm{n}^r \cdot \bm{s} \cdot \bm{n}^l \in \mathcal{N} \otimes \mathcal{S} \otimes \mathcal{N}$. Based on the definitions of transitive verb and noun, we can transform a sentence like ``Alice likes Bob" into the following form 
\begin{equation}
\bm{n} \cdot (\bm{n}^r \cdot \bm{s} \cdot \bm{n}^l) \cdot \bm{n} \rightarrow (\bm{n} \cdot \bm{n}^r) \cdot \bm{s} \cdot (\bm{n}^l \cdot \bm{n}) \rightarrow \bm{1} \cdot \bm{s} \cdot \bm{1} \rightarrow \bm{s}.
\end{equation}
The brackets in the above equation represent a process of dimension reduction/contraction, which is equivalent to
\begin{equation}
\sum_i \langle ii| \otimes \bm{s} \otimes \sum_i \langle ii|
\end{equation}
in Dirac notation.
The reduction operation in the above expression is realized through entanglement and measurement operations on qubits, i.e.,
\begin{equation}
\Qcircuit @C=1em @R=.7em {
& \ctrl{1} & \qw & \meter  \\
& \targ & \qw &  \qw
} 
\end{equation}

The above quantum operation in dimension reduction can also be realized classically using a matrix operation, i.e., tensor contraction.
A tensor after contraction can be another tensor or a matrix. For example, tensor $T_{ijk}\delta_{jk}=\sum_k T_{ijk} =T_{ijj} =M_{ij}$. A matrix after contraction can be a lower-rank matrix or a vector. For example, matrix $T_{ij} \delta_{ij}=T_{ii}$ is a vector. 
The quantum operations and tensor methods allow us to implement the QNLP ans\"atze in various ways.

\subsection{Ans\"atze and implementation}
In quantum physics, an ansatz is a variational formulation of a wave function that provides an exact or close solution to a Hamiltonian. In this study, an ansatz is a parametrized map with unknown parameters that transforms a string diagram into a specific quantum circuit. This transform maps each string to one qubit, and each word corresponds to a few parametrized quantum gates. The unknown parameters are determined by optimization with the given training corpora.
There are different ans\"atze used in QNLP, such as \texttt{IQPAnsatz} (Instantaneous Quantum Polynomial), \texttt{StronglyEntanglingAnsatz}, \texttt{Sim14Ansatz}, and \texttt{Sim15Ansatz}. \texttt{IQPAnsatz} primarily consists of Hadamard and rotation gates (e.g., $R_z$, $R_x$).  The IQP ansatz has been widely used in other contexts, such as a quantum kernel in support vector machines \cite{havlivcek2019supervised,pei2024designing}.
\texttt{StronglyEntanglingAnsatz} uses three single qubit rotations ($R_z$, $R_y$, $R_z$) followed by a ladder of CNOT gates to entangle the qubits strongly. \texttt{Sim14Ansatz} and \texttt{Sim15Ansatz} were proposed by Sim {\it et al.} and named after the first author \cite{sim2019expressibility}. They have similar structures based purely on rotation gates ($R_x$, $R_y$); the difference is that \texttt{Sim15Ansatz} has only half of the parameters as \texttt{Sim14Ansatz} and thus is considered less expressibility. The quantum circuits for these ans\"atze can be found in Ref. \cite{khatriexperimental} and not repeated here.

The lambeq package is the implementation of various ans\"atze to convert sentences into string diagrams according to pregroup grammar and eventually into quantum circuits that can be computed on quantum simulators or quantum hardware \cite{kartsaklis2021lambeq}. The implemented quantum circuit consists of initializations of qubit states, sequenced quantum gates to manipulate quantum states, and measurement operations to read the probabilistic states. 

Another mathematical realization is to use a tensor network to represent the ans\"atze. These include \texttt{MPSAnsatz} that is a tensor network-based approach leveraging Matrix Product States (MPS), \texttt{SpiderAnsatz} that is a representation inspired by categorical quantum mechanics (CQM), where structures are encoded in compact tensor diagrams, and \texttt{TensorAnsatz} that is a general tensor contraction framework that offers flexible encoding of linguistic structures.  These ans\"atze leverage tensor algebra to encode information efficiently. A quantum gate is mathematically analogous to a matrix: a single-qubit gate is represented as a 2$\times$2 matrix, while a two-qubit gate is a 4$\times$4 matrix. A tensor network is similar to a classical neural network. It is a directed acyclic graph to describe linear algebraic operations between tensors. 
The vertices of the graph are multi-linear tensor maps, and the edges correspond to vector spaces. 
The string diagrams can be viewed as a special tensor network.
It is worth mentioning that a tensor network is a classical realization of quantum circuits without involving complex numbers. It has been shown that although we can use two real numbers to represent complex numbers, complex numbers give higher fidelity in quantum information experiments \cite{chen2022ruling}.




\section{Experimental details}

This section describes the experimental setup designed to evaluate the performance of various ans\"atze and hyperparameters in text classification tasks. We will test the capability of QNLP models in classifying sentences into two groups under various circumstances. The primary focus of the experiments is to systematically analyze the influence of rewriting schemes, types of ans\"atze, and hyperparameter configurations on classification performance using a dataset known as \texttt{mc\_data} \cite{lorenz2023qnlp}. Each sentence in the dataset either belong to a ``IT" topic or ``food" topic, which can be taken as 0 or 1. This dataset contains labeled sentences used for training and testing QNLP models. 

The ans\"atze in this study are divided into two categories: circuit-based and tensor-based ans\"atze. Circuit-based ans\"atze represent quantum operations as quantum circuits composed of gates with specified parameters. Examples include \texttt{IQPAnsatz}, \texttt{StronglyEntanglingAnsatz}, \texttt{Sim14Ansatz}, and \texttt{Sim15Ansatz}. In contrast, tensor-based ans\"atze represent quantum gates as tensors, where a one-qubit gate corresponds to a $2 \times 2$ matrix, and a two-qubit gate corresponds to a $4 \times 4$ matrix. Tensor-based ans\"atze such as \texttt{MPSAnsatz}, \texttt{SpiderAnsatz}, and \texttt{TensorAnsatz} leverage this structure to perform efficient tensor contractions.

We employ the \texttt{Lambeq} framework \cite{kartsaklis2021lambeq} for constructing, rewriting, and training string diagrams. The rewriting operations applied to these diagrams, such as \texttt{re}, \texttt{re\_norm}, \texttt{re\_norm\_cur}, and \texttt{re\_norm\_cur\_norm}, simplify or modify the structure of diagrams to improve computational efficiency and accuracy. These rewriting schemes serve as a preprocessing step before constructing quantum circuits or tensors for text classification.

The experimental evaluation consists of four main parts:
\begin{itemize}
    \item \textbf{Rewriter Dependent Performance on Text Classification:} This experiment examines how different rewriting schemes affect the training and validation performance of QNLP models using the \texttt{IQPAnsatz}.
    \item \textbf{Performances of Circuit-based Ans\"atze:} This experiment compares the performance of various circuit-based ans\"atze by measuring their training and validation losses and accuracies across multiple configurations.
    \item \textbf{Circuit Hyperparameter Dependent Performances:} In this experiment, we investigate the impact of varying the number of layers (\texttt{n\_layers}) and single-qubit rotations (\texttt{n\_single\_qubit\_params}) on the training dynamics and final accuracy of QNLP models.
    \item \textbf{Performances of Tensor-based Ans\"atze:} The focus here is to analyze the performance of tensor-based ans\"atze, evaluating their generalization ability and their potential advantage in handling high-dimensional data representations.
\end{itemize}

By analyzing the results across these experiments, we aim to provide insights into the effects of rewriting schemes, ans\"atze types, and hyperparameter choices on the efficiency and accuracy of QNLP models. This study contributes to a better understanding of how different configurations influence the performance of quantum-based NLP systems and highlights optimal strategies for improving text classification accuracy.

\section{Results and discussion}

\subsection{Rewriter Dependent Performance on Text Classification}
\begin{figure*}[ht]
    \centering
    \includegraphics[width=1.0\linewidth]{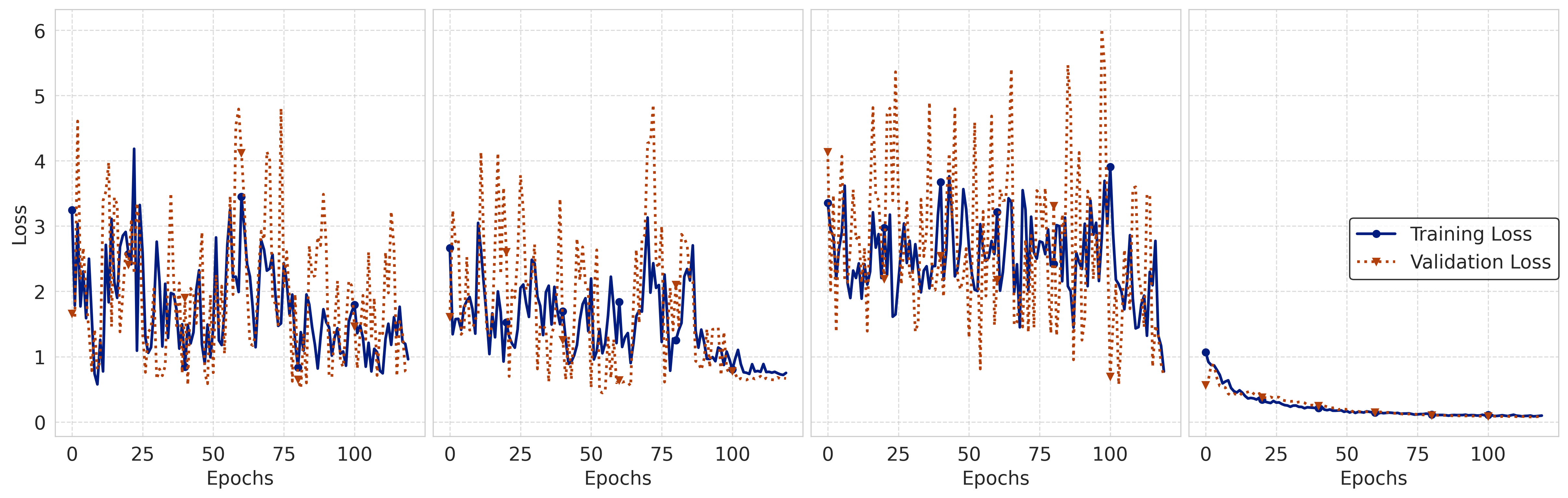}
    \caption{Loss curves for IQP ansatz with varying rewriters. Here we consider four schemes of rewriters, i.e., re, re\_norm, re\_norm\_cur, and re\_norm\_cur\_norm, respectively.}
    \label{fig:iqp-rewriters}
\end{figure*}

Rewriting schemes play a crucial role in the performance of quantum natural language processing models by determining the structure of the diagrams that represent linguistic inputs. These rewriters simplify, transform, or enrich the input diagrams, which in turn affects the expressivity of the quantum circuits derived from them. Figure~\ref{fig:iqp-rewriters} shows the impact of different rewriting schemes on the train and validation losses. Here, we take the 	
Instantaneous Quantum Polynomial (IQP) ansatz in a text classification task as an example.

\begin{itemize}
    \item \textbf{Raw Rewriter (\texttt{re})}: This scheme introduces minimal preprocessing, preserving the original complexity of the diagrams. The raw rewriter often leads to significant fluctuations in the training and validation losses, as shown in Figure~\ref{fig:iqp-rewriters}. The lack of simplification results in diagrams with high variability in structure, leading to inconsistent optimization and poor convergence. This is evident in the highly unstable loss curves for the \texttt{re} rewriter.

    \item \textbf{Normalized Rewriter (\texttt{re\_norm})}: By applying a normalization step, this rewriter simplifies the input diagrams further. It stretches the wires in the string diagram, and the consequence is that some qubits are combined. For example, the left and right adjoints (e.g., $\bm{n}^l$ and $\bm{n}$) share one qubit rather than use two separate qubits. Normalization standardizes the diagram structures, removing unnecessary complexity and redundancies. This results in smoother and more stable loss curves, as shown in Figure~\ref{fig:iqp-rewriters}. The standardization improves the convergence of the optimization process by reducing the variability in the quantum circuits.

    \item \textbf{Curry Rewriter After Normalization (\texttt{re\_norm\_cur})}: After normalization, the Curry transformation is applied, which introduces additional parameters to the diagrams by decomposing multi-argument functions into single-argument forms. This increases the complexity of the diagrams and consequently, the quantum circuits derived from them. While this increases the model's expressive power, it also reintroduces variability and fluctuations in the training and validation losses. This explains the less stable loss curves compared to \texttt{re\_norm}.

    \item \textbf{Normalized Rewriter After Currying (\texttt{re\_norm\_cur\_norm})}: The combination of normalization and currying followed by another normalization step results in the most stable and smooth loss curves. The final normalization reduces the variability introduced by the Curry rewriter, simplifying the diagram relationships without significantly reducing the expressivity of the model. This allows for effective optimization and leads to the best performance in terms of both training stability and validation loss reduction.
\end{itemize}

The progression of rewriting schemes demonstrates how the balance between simplification and expressivity affects model performance. The raw rewriter preserves the original structure at the cost of optimization stability, while the re\_norm\_cur\_norm rewriter achieves the best performance by iteratively refining the structure to balance expressivity and stability.

The rewriter-dependent results show that rewriting schemes significantly impact the performance of QNLP models. By systematically simplifying and enriching diagram structures, the re\_norm\_cur\_norm rewriter demonstrates the potential of effective preprocessing to improve the convergence and generalization of quantum NLP models. These findings underscore the importance of diagram optimization in developing robust QNLP systems.

\subsection{Performances of Circuit-based Ans\"atze}

The choice of ans\"atze plays a critical role in determining the performance of QNLP models, as it directly influences the expressivity, trainability, and generalization of the underlying quantum circuits. Different ans\"atze encode varying levels of complexity and parameterization, which interact uniquely with the structure of the input diagrams generated by the rewriter. In this subsection, using the MC dataset, we analyze the performance of four circuit-based ans\"atze—\texttt{IQPAnsatz}, \texttt{StronglyEntanglingAnsatz}, \texttt{Sim14Ansatz}, and \texttt{Sim15Ansatz}—on a text classification task. By comparing the mean training and validation losses and accuracies of the last ten epochs (Table~\ref{tab:loss-accuracy-varying-ansatzes}), we aim to uncover the strengths and limitations of each ansatz and identify the factors contributing to their performance differences. This comparison sheds light on how specific ansatz designs align with the input structures generated by the \texttt{re\_norm\_cur\_norm} rewriter, providing insights into optimizing QNLP pipelines.

\begin{table}[!htb]
    \centering
    \scriptsize
    \setlength{\tabcolsep}{5pt} 
    \renewcommand{\arraystretch}{1.2} 
    \begin{tabular}{lcccc}
        \toprule
         & \textbf{IQP} & \textbf{StronglyEntangling} & \textbf{Sim14} & \textbf{Sim15} \\
        \midrule
        \textbf{Train Loss} & 0.0916 & 0.2038 & 0.1247 & 0.1714 \\
        \textbf{Val Loss} & 0.0801 & 0.2284 & 0.0931 & 0.1824 \\
        \textbf{Train Accuracy} & 0.9986 & 0.9450 & 0.9879 & 0.9671 \\
        \textbf{Val Accuracy} & 0.9667 & 0.9333 & 1.0000 & 0.9700 \\
        \bottomrule
    \end{tabular}
    \caption{Mean loss and accuracy of the last ten epochs for classification tasks on the MC dataset with varying ans\"atze paired with re\_norm\_cur\_norm rewriter.}
    \label{tab:loss-accuracy-varying-ansatzes}
\end{table}

The results in Table~\ref{tab:loss-accuracy-varying-ansatzes} demonstrate the impact of different ans\"atze on the training and validation performance of the QNLP model using the \texttt{re\_norm\_cur\_norm} rewriter. Among the evaluated ans\"atze, \texttt{Sim14Ansatz} achieves a mean validation accuracy of 1.0 and maintains a low mean validation loss of 0.0931, indicating its robustness and superior generalization ability for this classification task. The \texttt{IQPAnsatz} also performs exceptionally well, with the highest mean training accuracy (0.9986), and a validation accuracy (0.9667), coupled with the lowest training and validation losses (0.0916 and 0.0801, respectively), suggesting that it is generally highly effective for the given dataset when paired with the \texttt{re\_norm\_cur\_norm} rewriter.

In contrast, \texttt{StronglyEntanglingAnsatz} and \texttt{Sim15Ansatz} exhibit slightly lower performance. While both achieve reasonable training and validation accuracies (e.g., 0.945 and 0.9333, respectively, for \texttt{StronglyEntanglingAnsatz}), their higher losses (0.2038 for training and 0.2284 for validation in \texttt{StronglyEntanglingAnsatz}) suggest that they are less efficient in capturing the patterns in the data compared to \texttt{Sim14Ansatz} and \texttt{IQPAnsatz}. The \texttt{Sim15Ansatz} shows a moderate trade-off, achieving relatively high validation accuracy (0.97) but with higher losses than \texttt{Sim14Ansatz}. This difference is not difficult to understand given the fact that \texttt{Sim15Ansatz} has a similar structure with \texttt{Sim14Ansatz} but its parameter size is only half of the latter. The less expressivity of \texttt{Sim15Ansatz} explains its lower accuracy. 

These results highlight the importance of selecting an appropriate ans\"atz to balance training efficiency, generalization capability, and circuit complexity. The superior performance of \texttt{Sim14Ansatz} and \texttt{IQPAnsatz} can be attributed to their optimal parameterization, which aligns well with the simplified input representations produced by the \texttt{re\_norm\_cur\_norm} rewriter.

\subsection{Circuit Hyperparameter Dependent Performances}

In this subsection, we examine the performance of ans\"atze in text classification tasks by varying its key hyperparameters: the number of layers (\texttt{n\_layers}) and the number of single-qubit rotation parameters (\texttt{n\_single\_qubit\_params}). The hyperparameter \texttt{n\_layers} controls the depth of the circuit, where additional layers introduce more entanglement and complexity by stacking parameterized gates sequentially. In contrast, \texttt{n\_single\_qubit\_params} defines the number of single-qubit operations, which directly affect the expressivity of the circuit by adjusting the individual rotations applied to each qubit. These parameters together shape the structure of the quantum circuit, influencing its ability to capture intricate patterns in the data.

Our analysis explores how different combinations of these hyperparameters impact the training dynamics, including the convergence behavior and generalization performance. For each combination of hyperparameters—\texttt{n\_layers} (circuit depth) and \texttt{n\_single\_qubit\_params} (number of single-qubit rotations)—we trained the model using the same preprocessed diagrams obtained through the \texttt{re\_norm\_cur\_norm} rewriting scheme. This rewriting method enhances the input data representation by normalizing and curving the compositional structure, ensuring a consistent and interpretable representation. This setup allows us to observe the effect of increasing circuit depth and rotation parameters on classification accuracy, particularly when applied to the MC dataset.

In this experimental setup, we evaluate the performance of ans\"atze with varying hyperparameter configurations on a dedicated test dataset, \texttt{mc\_test\_data}, which is a held-out portion of the data specifically reserved for assessing the generalization capabilities of the trained models. The evaluation was conducted after training the model for 120 epochs. During this process, we monitored the training and validation losses and accuracies to track the convergence behavior and ensure that the model was not overfitting or underfitting.

\begin{figure}[!htb]
    \centering
    \includegraphics[width=1.0\linewidth]{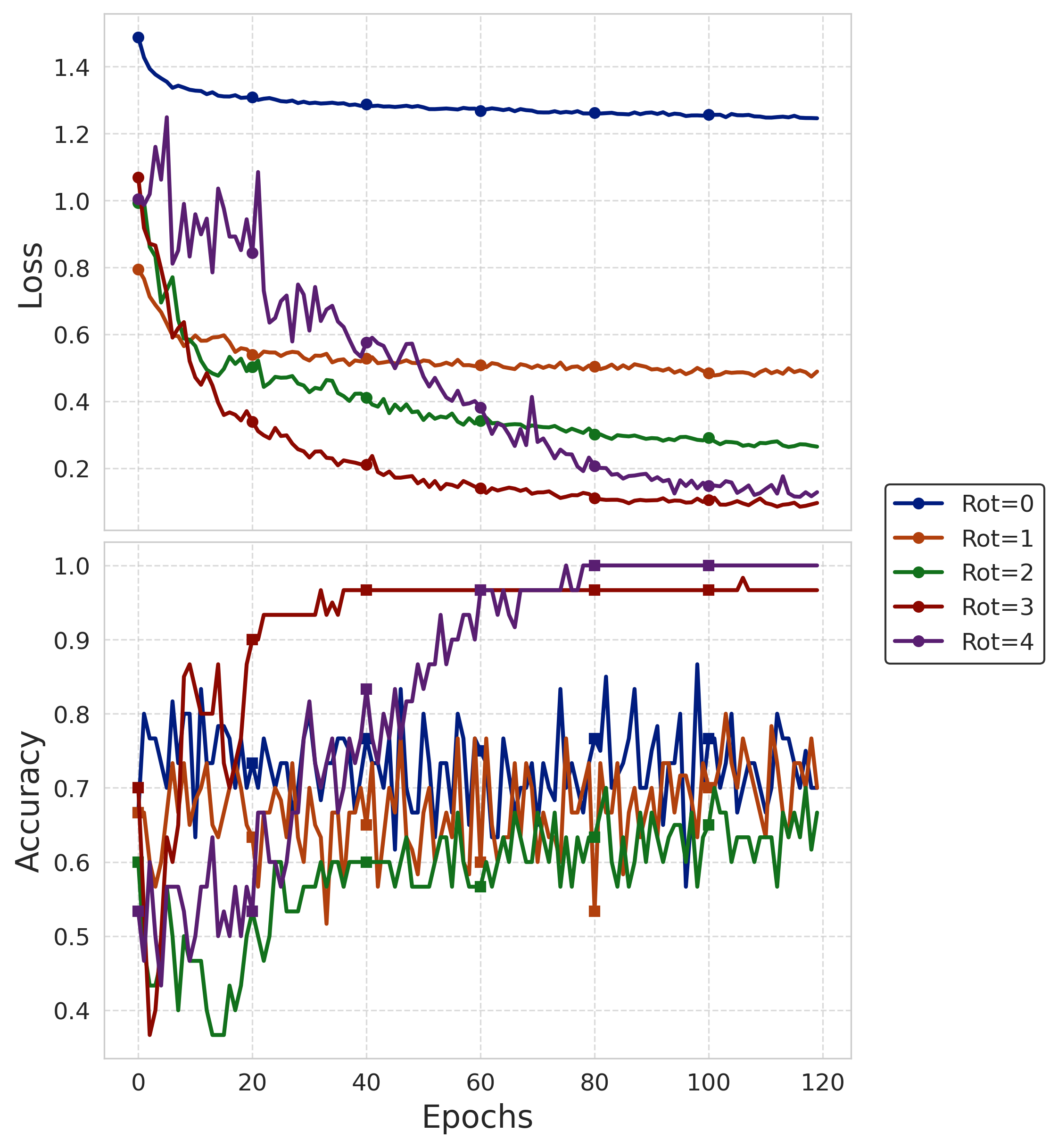}
    \caption{Training loss and validation accuracy across varying single-qubit rotations with fixed circuit depth value (\texttt{n\_layers=2}).}
    \label{fig:n_layers}
\end{figure}

The training and validation loss and accuracy for \texttt{IQPAnsatz} with an increasing number of single qubit rotations with a fixed number of \texttt{n\_layers=2} are shown in Figure~\ref{fig:n_layers}.
The results show that the configuration with \texttt{n\_layers=2} and \texttt{n\_single\_qubit\_params=3} demonstrates the fastest convergence, with the training loss stabilizing by around 40 epochs. Additionally, the combination of \texttt{n\_layers=2} and \texttt{n\_single\_qubit\_params=4} achieves a near-perfect validation accuracy of 1.0 at later epochs, indicating strong generalization performance. As the number of single-qubit rotations increases, the training loss tends to decrease more rapidly while the validation accuracy stabilizes more smoothly. This trend is particularly noticeable when comparing the fluctuations in the loss and accuracy curves for rotations 0 and 1, which show slower convergence, to those for rotations 3 and 4, which exhibit smoother and more stable behavior.

The number of rotations influences the stability and efficiency of the training process. Configurations with lower rotation values (such as 0 and 1) display slower convergence and occasional plateaus in the reduction of loss, suggesting potential underfitting. In contrast, higher rotation values (e.g., 3 and 4) lead to faster loss reduction and more consistent accuracy improvements over time. 

A key reason for this performance trend lies in the expressive power provided by the circuit hyperparameters. With \texttt{n\_layers=2}, the circuit maintains sufficient expressiveness without introducing excessive depth that could lead to optimization challenges or noise accumulation. This balance between simplicity and representational capacity facilitates faster convergence. Moreover, increasing the number of single-qubit rotations introduces additional degrees of freedom, enabling the ansatz to approximate more complex transformations effectively. Consequently, configurations with higher rotations fit the training data more efficiently, resulting in lower loss and higher validation accuracy.

Despite the increased parameterization for \texttt{n\_single\_qubit\_params=4}, the model avoids overfitting, likely due to the preprocessing step using the \texttt{re\_norm\_cur\_norm} rewriter. This rewriter simplifies the input diagrams, reducing unnecessary complexity and enhancing the model’s generalization capabilities. On the other hand, configurations with fewer rotations, such as 0 or 1, may have limited representational capacity, which explains the slower convergence and the suboptimal performance observed in the validation accuracy.

After completing the training phase, we evaluated the model’s performance on the \texttt{mc\_test\_data} to measure its ability to classify unseen data accurately. The results, shown in Table~\ref{tab:combined_comparison}, provide a comprehensive overview of the test accuracy across different configurations of \texttt{n\_layers} and \texttt{n\_single\_qubit\_params} for various ans\"atzes (IQP, Strongly Entangling, Sim14, and Sim15). These configurations were chosen to assess the trade-off between circuit complexity and classification performance. These findings provide insights into the balance between model complexity and performance, emphasizing the importance of careful hyperparameter tuning to achieve optimal outcomes.

\begin{table*}[!hbt]
    \centering
    \scriptsize
    \setlength{\tabcolsep}{6pt}
    \renewcommand{\arraystretch}{1.2}
    \begin{tabular}{cccccc@{\hskip 20pt}ccccc}
        \toprule
        & \multicolumn{5}{c}{\textbf{IQP}} & \multicolumn{5}{c}{\textbf{StronglyEntangling}} \\
        \cmidrule(lr){2-6} \cmidrule(lr){7-11} 
                            \textbf{\# Layers} & \rotatebox{90}{\textbf{0 Rot.}} & \rotatebox{90}{\textbf{1 Rot.}} & \rotatebox{90}{\textbf{2 Rot.}} & \rotatebox{90}{\textbf{3 Rot.}} & \rotatebox{90}{\textbf{4 Rot.}} & \rotatebox{90}{\textbf{0 Rot.}} & \rotatebox{90}{\textbf{1 Rot.}} & \rotatebox{90}{\textbf{2 Rot.}} & \rotatebox{90}{\textbf{3 Rot.}} & \rotatebox{90}{\textbf{4 Rot.}} \\
        \midrule
        0                &NaN&0.367&0.533&0.533&0.400     &NaN&0.500&0.417&0.400&0.500    \\
        1                &0.767&\textbf{0.967}&0.567&0.833&\textbf{0.967}   &0.800&0.800&0.633&\textbf{1.000}&0.900    \\
        2                &0.817&0.867&0.833&\textbf{0.967}&\textbf{1.000}   &0.767&0.700&0.600&\textbf{0.933}&\textbf{1.000}    \\
        3                &0.767&0.467&0.900&\textbf{0.967}&\textbf{0.933}   &0.767&0.800&0.433&0.767&0.800    \\
        4                &0.700&0.667&0.833&0.900&\textbf{0.967}   &0.767&0.733&\textbf{0.967}&0.700&0.667    \\
        \bottomrule
    \end{tabular}
    \begin{tabular}{cccccc@{\hskip 20pt}ccccc}
        & \multicolumn{5}{c}{\textbf{Sim14}} & \multicolumn{5}{c}{\textbf{Sim15}} \\
        \cmidrule(lr){2-6} \cmidrule(lr){7-11} 
                            \textbf{\# Layers} & \rotatebox{90}{\textbf{0 Rot.}} & \rotatebox{90}{\textbf{1 Rot.}} & \rotatebox{90}{\textbf{2 Rot.}} & \rotatebox{90}{\textbf{3 Rot.}} & \rotatebox{90}{\textbf{4 Rot.}} & \rotatebox{90}{\textbf{0 Rot.}} & \rotatebox{90}{\textbf{1 Rot.}} & \rotatebox{90}{\textbf{2 Rot.}} & \rotatebox{90}{\textbf{3 Rot.}} & \rotatebox{90}{\textbf{4 Rot.}} \\
        \midrule
        0                &NaN&0.367&0.533&0.533&0.400     &NaN&0.367&0.533&0.533&0.400    \\
        1                &0.833&\textbf{0.933}&0.817&0.700&\textbf{0.967}   &0.767&0.800&0.900&0.867&\textbf{0.967}    \\
        2                &0.767&0.467&0.833&\textbf{1.000}&0.900   &0.833&\textbf{0.933}&\textbf{0.967}&\textbf{1.000}&\textbf{0.967}    \\
        3                &0.733&0.833&0.800&0.600&0.667   &0.767&\textbf{0.967}&0.700&\textbf{0.967}&0.867    \\
        4                &0.767&0.767&0.733&0.500&0.467   &0.767&0.717&0.683&\textbf{0.967}&0.900    \\
        \bottomrule
    \end{tabular}
    \caption{Comparison of test accuracy of four ans\"atze with varying hyperparameters. More specifically, we consider re\_norm\_cur\_norm rewriter across ans\"atze, number of layers (depth), and number of single-qubit rotations. ``NaN" represents no results available.}
    \label{tab:combined_comparison}
\end{table*}

The table highlights the impact of increasing the number of layers and single-qubit rotations. For instance, adding more layers generally increases the expressiveness of the model but at the cost of computational complexity and the risk of overfitting. Similarly, the number of single-qubit rotations determines the diversity of local operations applied to each qubit, which can either improve or degrade the model’s performance based on the dataset and the overall circuit design. The results illustrate that optimal performance is achieved when the circuit strikes a balance between depth and parameterization, as seen in specific combinations such as \texttt{n\_layers = 2} and \texttt{n\_single\_qubit\_params = 3}.

\subsection{Performances of Tensor-based Ans\"atze}

To assess the effectiveness of tensor-based ansatz, we conducted a text classification experiment using the MC dataset, where sentences were first processed through either the re or re\_norm. The models were trained for 120 epochs, tracking training and validation loss and accuracy metrics.
The tensor-based ansatz evaluated include \texttt{MPSAnsatz}, \texttt{SpiderAnsatz}, and \texttt{TensorAnsatz}.
Each ansatz was trained using the \texttt{PytorchModel} backend, which allows efficient tensor computations, in contrast, circuit-based ansatz utilizes \texttt{TketModel} or \texttt{PennyLaneModel} for quantum circuit compilation and execution.


The results presented in Table~\ref{tab:tensor-performance} highlight the performance of different tensor-based ans\"atze—MPS, Spider, and Tensor—when paired with the \texttt{re} and \texttt{re\_norm} rewriters. A key observation is that all models achieve a training accuracy of 1.0, indicating their ability to learn the training data entirely. However, differences emerge in their generalization to validation data, as reflected in the validation loss and accuracy values. Among the three ans\"atze, the \texttt{SpiderAnsatz} achieves the best validation performance with the \texttt{re} rewriter, reaching a validation accuracy of 1.0 and maintaining the lowest validation loss (0.0064).

\begin{table}[!hbt]
    \centering
    \scriptsize
    \setlength{\tabcolsep}{6pt}
    \renewcommand{\arraystretch}{1.2}
    \begin{tabular}{ccccccc}
        \toprule
        & \multicolumn{2}{c}{\textbf{MPS}} & \multicolumn{2}{c}{\textbf{Spider}} & \multicolumn{2}{c}{\textbf{Tensor}}\\
        \cmidrule(lr){2-3} \cmidrule(lr){4-5} \cmidrule(lr){6-7} 
                             & \textbf{re} & \textbf{re\_norm} & \textbf{re} & \textbf{re\_norm} & \textbf{re} & \textbf{re\_norm} \\
        \midrule
        \textbf{Train Loss}             &0.0003&0.0008&0.0002&0.0006&0.0003&0.0007    \\
        \textbf{Val Loss}            &0.0723&0.0897&0.0064&0.0208&0.0723&0.0897    \\
        \textbf{Train Acc.}      &1.0&1.0&1.0&1.0&1.0&1.0    \\
        \textbf{Val Acc.}          &0.9833&0.9833&1.0&0.9833&0.9833&0.9833    \\
        \bottomrule
    \end{tabular}
    \caption{Mean loss and accuracy of the last ten epochs for classification tasks on the MC dataset with varying tensor-based ans\"atze and rewriter combinations.}
    \label{tab:tensor-performance}
\end{table}
 
This suggests that \texttt{SpiderAnsatz}, when paired with \texttt{re}, generalizes exceptionally well to unseen data. The other two ans\"atze, MPS and Tensor, achieve slightly lower validation accuracy (0.9833) and exhibit higher validation loss values, particularly when trained with the \texttt{re\_norm} rewriter, which slightly increases the validation loss across all models. This indicates that the additional normalization step in \texttt{re\_norm} might introduce some form of regularization that prevents overfitting but also marginally reduces generalization performance.

\begin{figure*}[!htb]
    \centering
    \includegraphics[width=\linewidth]{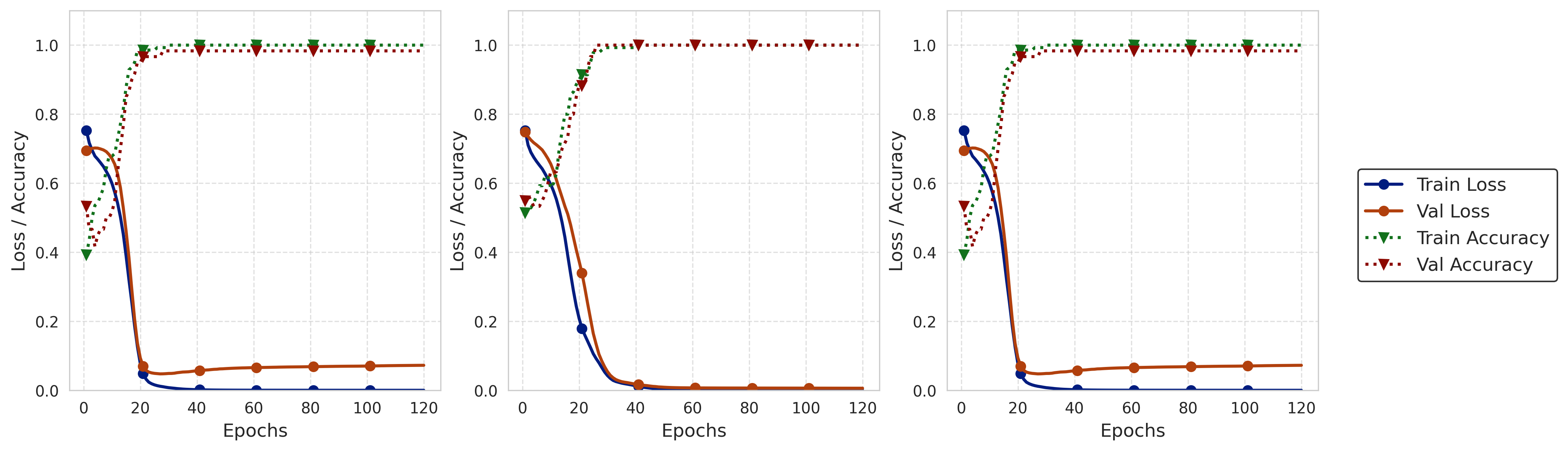}
    \caption{Loss and accuracy of training using raw rewriter (\texttt{re}), with MPS, Spider, and Tensor ansatz, respectively.}
    \label{fig:tensor-ansatz-re}
\end{figure*}

\begin{figure*}[!htb]
    \centering
    \includegraphics[width=\linewidth]{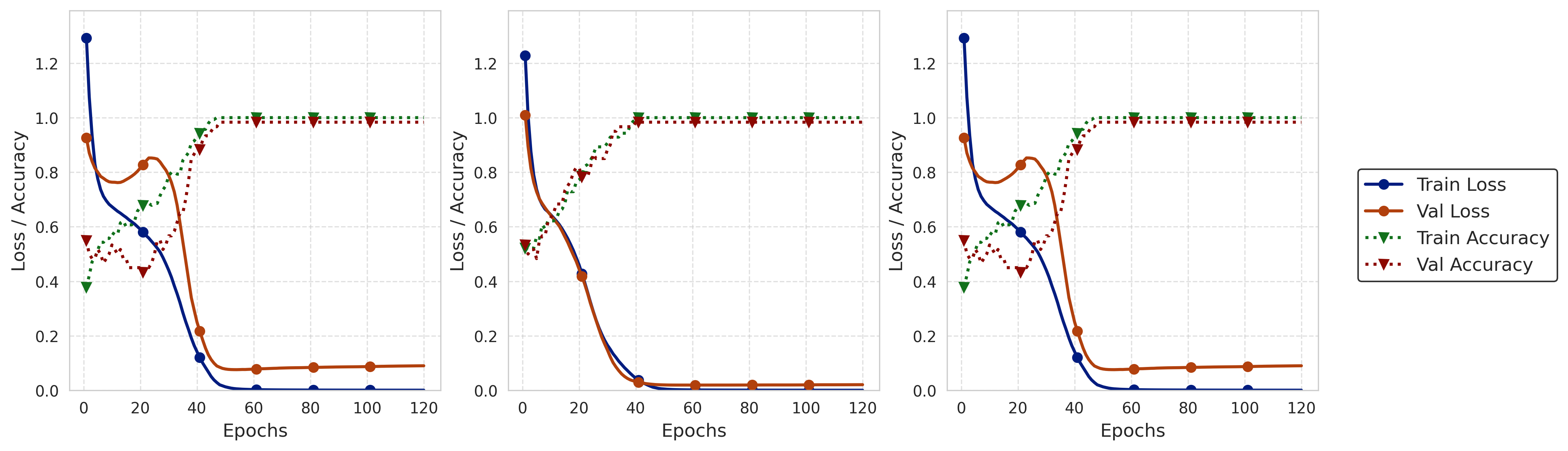}
    \caption{Loss and accuracy of training using raw rewriter and normalization (\texttt{re\_norm}), with MPS, Spider, and Tensor ansatz, respectively.}
    \label{fig:tensor-ansatz-re-norm}
\end{figure*}

The performance of tensor-based ansatze (\texttt{MPSAnsatz}, \texttt{SpiderAnsatz}, and \texttt{TensorAnsatz}) using the \texttt{re} and \texttt{re\_norm} rewriters is visualized in Figures~\ref{fig:tensor-ansatz-re} and~\ref{fig:tensor-ansatz-re-norm}. The results indicate key differences in convergence speed, generalization, and overfitting behavior across different configurations.

The speed of convergence varies depending on the rewriter used. Models trained with the \texttt{re} rewriter reach a validation accuracy of 1.0 within approximately 30 epochs, whereas those using the \texttt{re\_norm} rewriter require around 50 epochs. This suggests that the additional normalization step in \texttt{re\_norm} smooths the training process but slightly delays convergence. The loss curves in Figure~\ref{fig:tensor-ansatz-re} show sharper decreases in loss when using \texttt{re}, while Figure~\ref{fig:tensor-ansatz-re-norm} demonstrates a more gradual and controlled descent in loss values.

Overfitting patterns differ between the two rewriters. For all three ansatze, models trained with the \texttt{re} rewriter begin to exhibit overfitting around epoch 40, where the training loss continues to decrease while the validation loss plateaus or slightly increases. In contrast, models trained with \texttt{re\_norm} start showing overfitting later, at approximately epoch 60. This suggests that the normalization step in \texttt{re\_norm} introduces a regularization effect that delays overfitting and promotes better generalization in the earlier training phase.

Additionally, the training curves indicate that \texttt{re\_norm} results in slightly higher validation loss in later epochs compared to \texttt{re}, which may be attributed to the added complexity of normalization modifying the quantum circuit representations. The slightly delayed but more stable convergence suggests that \texttt{re\_norm} may be preferable in scenarios where robustness against overfitting is prioritized.

Across both rewriters, the three tensor-based models follow nearly identical training trajectories, with only minor variations in the rate of initial loss reduction. The \texttt{SpiderAnsatz} demonstrates slightly more stable loss reduction during the training stages compared to the other two ansatze, which is consistent in both Figures~\ref{fig:tensor-ansatz-re} and~\ref{fig:tensor-ansatz-re-norm}. However, all three ans\"atze ultimately reach comparable final performance, achieving near-perfect classification accuracy by the later epochs.

Overall, these results reinforce the trends observed in previous experiments: while all tensor-based ans\"atze demonstrate strong classification capabilities, the choice of rewriter influences the rate of convergence and final validation performance. The \texttt{re} rewriter consistently enables faster convergence with lower validation loss, making it a suitable choice for rapid optimization. Meanwhile, \texttt{re\_norm} introduces a regularization effect that prevents early overfitting at the cost of slightly slower convergence. These findings suggest that selecting the appropriate rewriter should depend on the desired trade-off between convergence speed and model stability.

\subsection{Comparison of Circuit-based and Tensor-based Ans\"atze}

A key distinction in quantum NLP models lies in the representation of quantum gates and operations. Circuit-based ans\"atze are constructed using parameterized quantum circuits, where each quantum gate corresponds to a transformation applied to qubits. These ans\"atze are defined by a series of quantum gates that manipulate the quantum state. The expressiveness of these models depends on hyperparameters such as circuit depth (number of layers) and number of single-qubit parameters, both of which influence how well the model can capture complex relationships in data.
In contrast, tensor-based ans\"atze treat quantum computations as tensor network contractions rather than explicit gate-based operations.  In a tensor-based representation, the entire computation is formulated as a contraction of multiple tensors, bypassing the need for sequential circuit execution. This allows for efficient simulation on classical hardware while maintaining compatibility with quantum-inspired models.
The fundamental advantage of tensor-based approaches is that they can directly encode large-scale entanglement structures without the overhead of explicit quantum circuit compilation. This makes them particularly well-suited for classical simulations of quantum NLP models, as they can scale more efficiently when working with large datasets.

This study evaluated the performance of four circuit-based and three tensor-based quantum models. Overall, the performance of the tensor-based models is comparable to that of the circuit-based ones, albeit the former being slightly better than the latter. Nonetheless, the convergence behavior of the two groups of models is distinct. The convergence of the tensor-based models is more stable. It does not suffer from fluctuation, resulting from only using real numbers. Differently, the quantum behavior based on complex numbers has an outstanding probabilistic feature, resulting in significant model training fluctuations.









\section{Conclusions}

We have studied the performance of quantum natural language processing models at different levels in text classification tasks. We consider the influence of different schemes to create a string diagram, different ans\"atze to implement the diagram, and the hyperparameters in individual ans\"atze (e.g., the depth of quantum circuits and the number of single-qubit parameters). The rewriter-dependent results show that rewriting schemes impact the performance of quantum language models substantially. By systematically simplifying and enriching diagram structures, one of the four schemes (i.e., the \texttt{re\_norm\_cur\_norm} rewriter) demonstrates the potential of effective preprocessing to improve the convergence and generalization of quantum language models. These findings underscore the importance of diagram optimization in developing robust models. 
The results of different ans\"atze highlight the importance of selecting an appropriate ans\"atz to balance training efficiency, generalization capability, and circuit complexity.
We examine the performance of ans\"atze by varying its key hyperparameters.
Our results demonstrate how the balance between simplification and expressivity affects model performance.
Lastly, we have compared the above quantum version of ans\"atze based on quantum circuits and the classical version of quantum ans\"atze based on tensor operations. The performance of the tensor-based models is comparable to that of the circuit-based ones, but its convergence is more stable and does not suffer from fluctuation.
Our study identifies a few rules that make a high-performance quantum language models and paves the way toward designing new optimal quantum circuits for natural languages.


















\end{document}